\documentclass[11pt]{article}

\usepackage[preprint]{acl}
\usepackage{amsmath}
\usepackage{times}
\usepackage{latexsym}
\usepackage{booktabs}
\usepackage{pifont}
\usepackage{threeparttable}
\newcommand{\cmark}{\ding{51}}
\newcommand{\xmark}{\ding{55}}
\usepackage[T1]{fontenc}
\usepackage{multirow}

\usepackage[utf8]{inputenc}

\usepackage{microtype}

\usepackage{inconsolata}

\usepackage{graphicx}
\usepackage{amsmath,amssymb}
\title{ASR-Protected Fusion and Reference-Anchored Policy Optimization for Joint ASR and Word-Level Speech Anti-Spoofing}

\author{
  \textbf{Xueping Zhang\textsuperscript{2}},
  \textbf{Xiaoxiao Miao\textsuperscript{2}},
  \textbf{Tiantian Feng\textsuperscript{1}},
  \textbf{Ming Li\textsuperscript{1,*}}
  \\
  \textsuperscript{1}School of Artificial Intelligence, The Chinese University of Hong Kong, Shenzhen, China
  \\
  \textsuperscript{2}Digital Innovation Research Center, Duke Kunshan University, China
  \\
  \textsuperscript{*}Corresponding author: Ming Li, \href{mingli369@cuhk.edu.cn}{mingli369@cuhk.edu.cn}
}

\begin{document}
\maketitle
\begin{abstract}
Existing anti-spoofing systems report either an utterance-level score or a time-based spoof interval: the former cannot show which words are spoofed, while the latter does not directly match the transcript and must be post-aligned with words, so alignment errors can lead to wrong word-level labels. Standard ASR model outputs only text; spoof information in the audio is lost after recognition and is not passed with the transcript. Yet changing only a few words can change the meaning of a sentence, leaving downstream systems to use incorrect text without knowing which words are spoofed. We therefore treat partial anti-spoofing as tagged ASR: the model transcribes speech and marks spoofed words at the same time, directly linking spoof labels to output words.
We adopt the audio LLM and use ASR-Protected Subspace Gated Fusion (APSGF) to fuse spoof cues with ASR features. Spoof cues are added only outside an ASR-protected subspace, so they do not disturb recognition features. 
After cross-entropy training, we fine-tune the model with Reference-Anchored Policy Optimization (RAPO). Since the SFT policy is already concentrated, multiple samples from the same input tend to receive similar rewards, resulting in weak group-wise advantages. RAPO instead generates a single sample with low sampling randomness and anchors its advantage to the reward of a reference model. This allows the model to reinforce responses that outperform the reference while suppressing those that underperform it.
Experiments on five datasets demonstrate the effectiveness of our approach in jointly detecting spoofed words and preserving ASR accuracy. The code is publicly available \footnote{\url{https://anonymous.4open.science/r/asr_word_antispoofing_anon-38C6/}}.
\end{abstract}

\section{Introduction}

Speech deepfakes refer to synthetic or manipulated speech
\cite{sd,asv}
. Such manipulation may affect an entire utterance or only a small portion of it \cite{partialspoof, PartialEdit}. In partially spoofed speech, an attacker can replace a few critical words while leaving the surrounding speech intact, potentially changing the meaning of an utterance while retaining much of its original acoustic context. An utterance-level detector can flag the presence of manipulation, but cannot identify which words are unreliable. This motivates fine-grained localization of manipulated
content.
Existing approaches localize manipulated speech through frame-level classification\cite{cai2024integrating}, the detection of inconsistencies between neighboring frames, or manipulation-boundary detection \citet{zhang2026can}. These methods produce predictions along the audio timeline. To identify unreliable words in a transcript, their temporal predictions must be mapped to recognized words using word-level timestamps. Errors in transcription or word-boundary estimation can affect the resulting word-level spoof labels. Jointly generating a transcript and its spoof labels offers a way to associate authenticity predictions directly with the recognized content.

Recent studies jointly generate ASR transcripts and word-level spoof labels \cite{tran26_interspeech,zhang2026can}. This tagged ASR formulation attaches spoof predictions directly to recognized words, avoiding a separate temporal-to-word mapping at inference. It also carries authenticity information into the transcript, allowing downstream systems to identify potentially manipulated content when using the recognized text. However, joint generation alone does not ensure accurate transcription or reliable localization. \citet{zhang2026can}  report transcription difficulties in some settings and sensitivity to dataset-specific editing patterns. We therefore investigate two questions: how can spoof-related acoustic features be incorporated while preserving representations useful for ASR, and how can the feedback on both tasks guide the optimization of their joint output?

The first question concerns feature fusion. ASR emphasizes linguistic and phonetic information, whereas spoof detection also requires acoustic evidence of synthesis or manipulation. An auxiliary spoof encoder can provide complementary cues, but directly mixing its features with pretrained ASR representations may interfere with information used for transcription. We propose ASR-Protected Subspace Gated Fusion (APSGF) to constrain this interaction. APSGF constructs a protected subspace from the leading singular directions of bona fide ASR features and restricts gated fusion to its orthogonal complement. This preserves the original feature within the selected subspace while allowing spoof-related information to enter through the remaining directions.

The second question concerns joint optimization. After supervised fine-tuning with token-level cross-entropy, we use a sequence-level reward that evaluates both transcription and spoof-label errors to guide further optimization. Group-based methods can provide weak learning signals when low-temperature samples receive similar rewards, while increasing sampling randomness may lead to undesirable policy updates. We propose Reference-Anchored Policy Optimization (RAPO), which samples one response per input and compares its reward with that of a greedy response from a frozen post-SFT model. The advantage reinforces responses that outperform the reference and suppresses those that underperform it, while KL regularization constrains policy drift. This provides joint reward guidance without relying on within-group reward differences.

We evaluate the framework on five public English speech datasets: VCTK\cite{vctk}, PartialEdit\cite{PartialEdit}, AV-Deepfake1M \cite{AV}, SCDF \cite{Scdf}, and LlamaPartialSpoof \cite{llamapartialspoof}. These datasets cover bona fide speech, partially manipulated speech, and fully synthetic speech, with LlamaPartialSpoof reserved for out-of-domain evaluation. We consider a two-task setting that jointly generates transcripts and word-level spoof tags, and a multi-task setting that additionally predicts the utterance-level spoof scope. The complete framework improves average ASR and word-level detection performance over the corresponding SFT baselines. Ablation studies examine the contributions of feature fusion and policy optimization under both
in-domain and out-of-domain conditions. Our contributions are summarized as follows:
\begin{itemize}
\item We propose APSGF, which incorporates complementary spoof features through gated fusion in the orthogonal complement of a protected ASR feature subspace.
\item We propose RAPO, which uses a frozen reference reward to guide transcription and spoof localization without group-wise sampling.
\item We evaluate both components across five datasets and analyze their effects on transcription, word-level spoof detection, and out-of-domain generalization.
\end{itemize}

\section{Related work}
We review related work from three perspectives: speech spoof localization and its integration with ASR, feature fusion in audio language models, and reward-guided sequence policy optimization. 

\subsection{Word-level anti-spoofing and ASR}
Research on partially spoofed speech has extended detection from utterance-level decisions to fine-grained localization. PartialSpoof \cite{partialspoof} studies manipulated segments embedded in genuine speech, with frame- or segment-level predictions. HoliAntiSpoof \cite{xu2026holiantispoof} generates information about spoof presence, type, temporal location, and semantic impact. \citet{xue2026unifying} formulate speech-editing analysis as structured generation of edit types and edited content. These outputs are not necessarily transcripts with authenticity labels attached to individual words.

\citet{tran26_interspeech}  fine-tune Whisper\cite{whisper} to generate transcriptions and fake-word markers, while \citet{zhang2026can} investigate speech LLMs that produce transcripts with word-level spoof tags. Such outputs avoid a separate timestamp-to-word mapping at inference. However, \citet{zhang2026can} also report transcription difficulties and sensitivity to dataset-specific editing patterns. We build on tagged ASR to investigate complementary spoof features and joint reward feedback.

\subsection{Feature Fusion and Subspace Constraints}
Audio language models often combine multiple encoders to capture complementary properties of speech. SALMONN \cite{tang2024salmonn} combines Whisper and BEATs\cite{BEATs} features through a window-level Q-Former, while WavLLM \cite{2024wavllm} uses Whisper and WavLM\cite{wavlm} to incorporate linguistic and speaker-related information. Other approaches adapt the encoder combination to the input or prompt. MoWE-Audio \cite{zhang2025mowe} augments a base encoder with selected lightweight encoders, and PaM \cite{2025enhancing} uses a prompt-aware mixture of audio encoders. These methods illustrate different strategies for selecting and integrating complementary representations.

Gated fusion controls the contribution of different representations. \citet{gao2024speech} combine acoustic and ASR-derived semantic features using cross-attention and a temporal-aware gate for speech emotion recognition. \citet{wang25u_interspeech} apply gated fusion to Whisper encoder-layer features for multi-task speech learning. 
Subspace constraints and orthogonal projection can help preserve existing representations. GPM \cite{gpm} protects important activation subspaces by constraining parameter updates to their orthogonal complements, while SOP \cite{SOP} restricts feature modifications to orthogonal directions to preserve visual context. In contrast, APSGF constrains spoof-feature fusion to the orthogonal complement of selected ASR representation directions.

\subsection{Sequence Policy Optimization}
Regardless of the pipeline, RL was found to be an essential step in improving LLM performance across tasks and domains \cite{guo2025deepseek} and better preserving the prior knowledge acquired during pre-training \cite{shenfeld2026rl}.
In RL, sequence generation can be optimized using several forms of feedback beyond token-level supervision. DPO \cite{rafailov2023direct} learns from preferred and rejected responses, while SimPO \cite{meng2024simpo} uses length-normalized sequence
log-probabilities in a reference-free preference objective. PPO \cite{ppo}
estimates advantages using a learned value function. GRPO \cite{shao2024deepseekmath} instead uses relative rewards among responses sampled for the same input, and DAPO \cite{yu2026dapo} develops this approach with decoupled clipping and dynamic sampling, differing in both their feedback and baseline construction.

On-policy distillation \cite{opd} provides token-level teacher guidance on responses sampled from the current policy. ReMax \cite{remax} uses the reward of the current policy's greedy response as a baseline for a sampled response, without group-wise sampling. RAPO instead uses the reward of a low-stochasticity response from a frozen post-SFT model, computed once before optimization. It evaluates sampled responses against a fixed reference rather than a baseline that changes with the current policy.


\begin{figure}[t]
    \centering
    \includegraphics[width=\linewidth]{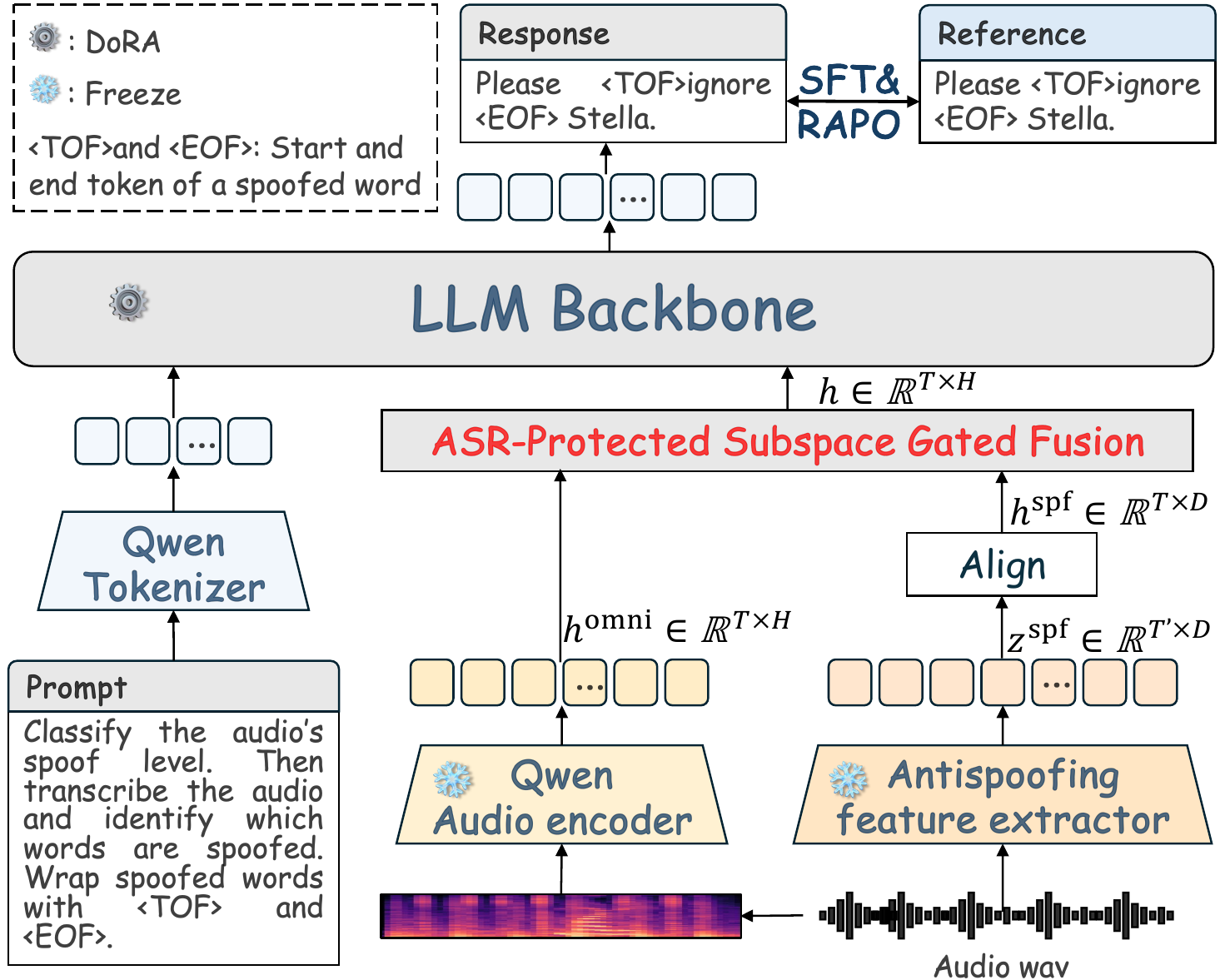}
    \caption{Overview of our framework for joint ASR and word-level spoof detection. ASR and spoof features from two frozen audio encoders are aligned and fused by APSGF, followed by unified transcript generation with spoof tags.}
    \label{fig:structure}
\end{figure}

\section{Method}
\subsection{Overall Framework Structure}
As illustrated in Figure~\ref{fig:structure}, our framework combines pretrained ASR features with auxiliary spoof-related features through APSGF, generates a transcript with word-level spoof tags, and is trained through supervised fine-tuning followed by RAPO.

\paragraph{Task Formulation.}
As shown in Figure ~\ref{fig:structure}, given a speech waveform $x \in \mathbb{R}^{N}$, our goal is to generate its transcript while identifying spoofed words. The model produces a token sequence $y=(y_1,\ldots,y_L)$.
Every detected spoofed word is enclosed by \texttt{<TOF>} and \texttt{<EOF>}, while a bona fide word is generated without these tags. Thus, transcription and word-level spoof detection are formulated as a unified sequence generation task.

\paragraph{Audio Feature Extraction.}

The input waveform $x$ is processed by two audio encoders: the Qwen2.5-Omni \cite{qwen25omni} audio encoder $f_{\mathrm{omni}}(\cdot)$ and a spoof encoder $f_{\mathrm{spf}}(\cdot)$. The spoof encoder $f_{\mathrm{spf}}(\cdot)$ is pretrained on the PartialSpoof dataset by the MultiReso \cite{partialspoof} method; it can extract frame-level spoof cues from the input audio waveform. The input waveform is converted into a log-mel spectrogram before being fed into the Qwen2.5-Omni audio encoder, while the spoof encoder directly takes the input waveform.
\begin{equation}
h^{\mathrm{omni}}
=
f_{\mathrm{omni}}\left(\operatorname{Mel}(x)\right)
\in
\mathbb{R}^{T\times H},
\end{equation}
\begin{equation}
z^{\mathrm{spf}}
=
f_{\mathrm{spf}}(x)
\in
\mathbb{R}^{T'\times D}.
\end{equation}
The Qwen2.5-Omni audio encoder provides the pretrained audio representation for the LLM, while the spoof encoder provides complementary spoof-related acoustic information. Here, $T$ and $T'$ denote the temporal lengths of the two feature sequences, while $H$ and $D$ denote their feature dimensions. Since the two encoders use different temporal resolutions, we align $z^{\mathrm{spf}}$ to the Omni audio-token positions as follows:
\begin{equation}
h^{\mathrm{spf}}
=
\operatorname{Align}
\left(z^{\mathrm{spf}}\right)
\in
\mathbb{R}^{T\times D},
\end{equation}
This alignment only matches the temporal resolutions of the two feature streams; it does not require word timestamps or map spoof predictions to transcript words.
We divide $z_{\mathrm{spf}}$ into $N$ temporal segments, where $N$ is the number of Qwen2.5-Omni audio tokens. We then use integer-truncated boundaries and average each segment to obtain token-aligned features $h_{\mathrm{spf}}$.

\begin{figure}[t]
    \centering
    \includegraphics[width=\linewidth]{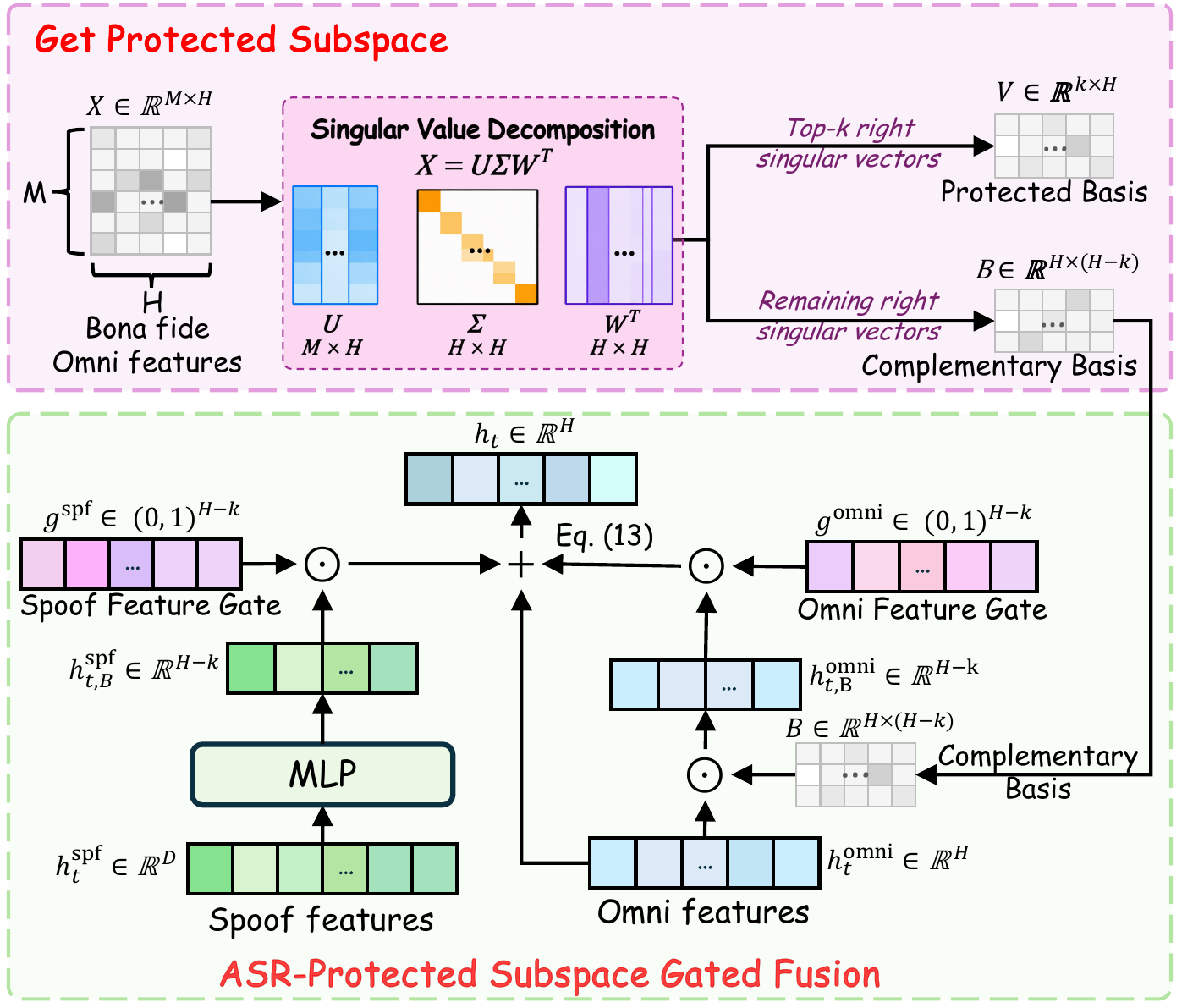}
    \caption{Overview of ASR-Protected Subspace Gated Fusion (APSGF). It first constructs a protected subspace and then performs gated fusion in the complementary subspace. $\odot$ denotes the Hadamard product, and $+$ denotes element-wise addition.}
    \label{fig:main2}
\end{figure}

\paragraph{Feature Fusion and Generation.}
We fuse the aligned features using APSGF as defined in Eq. ~\eqref{APSGF}. The fusion mechanism is detailed in Section~\ref{sec:apsgf}.
\begin{equation}
h
=
\mathrm{APSGF}
\left(
h^{\mathrm{omni}},
h^{\mathrm{spf}}
\right)
\in
\mathbb{R}^{T\times H}.
\label{APSGF}
\end{equation}

We then feed the fused representation into the language model to generate the tagged transcript. Given $h$ and a text prompt $\mathbf{p}$, the output sequence $y=(y_1,\ldots,y_L)$ is generated as follows:
\begin{equation}
p(y\mid h,\mathbf{p})
=
\prod_{t=1}^{L}
p(y_t\mid y_{<t},h,\mathbf{p}),
\label{tag}
\end{equation}
where $y_t$ denotes the token generated at step $t$. The output sequence contains both the transcript and word-level spoof tags.

The model is trained in two stages. We first perform Supervised Fine-Tuning (SFT) using the target transcription and spoof tags. We then apply Reference-Anchored Policy Optimization (RAPO) to further optimize the generation behavior. The detailed design of RAPO is described in Section~\ref {sec:rapo}.

\subsection{ASR-Protected Subspace Gated Fusion}
\label{sec:apsgf}

ASR-Protected Subspace Gated Fusion (APSGF) first constructs a protected subspace $V$ in the Omni representation and then injects spoof information only into its remaining orthogonal complement $B$. Because $VB = \mathbf{0}$, fusion leaves the projection onto the selected protected subspace unchanged. 
For simplicity, we describe APSGF for a single temporal position; the same operation is applied independently to all \(T\) positions.

\paragraph{Get Protected Subspace.}
As shown in the upper part of Figure~\ref{fig:main2}, given a set of bona fide utterances from the VCTK\cite{vctk} training set, we extract their audio features from the Omni audio encoder and collect the resulting feature vectors into a matrix $X\in\mathbb{R}^{M\times H}$, where $M$ is the number of collected feature vectors and $H$ is the feature dimension. We compute the Singular Value Decomposition as follows:
\begin{equation}
X=U\Sigma W^{\top},
\end{equation}
where $\Sigma$ contains the singular values $\sigma_1\geq\cdots\geq\sigma_H\geq0$, and the columns of $W$ are the corresponding right singular vectors. We define the protected basis $V$ as follows:
\begin{equation}
V = W_{[:,\, :k]}^\top \in \mathbb{R}^{k\times H},
\end{equation}
which consists of the first $k$ right singular vectors and captures the largest variance in the bona fide Omni features. We then compute an orthonormal basis of its orthogonal complement as follows:
\begin{equation}
B\in\mathbb{R}^{H\times(H-k)}.
\end{equation}
Thus, the subspace spanned by the rows of $V$ is the protected subspace, while the subspace spanned by the columns of $B$ contains the remaining directions where spoof information can be injected.

\paragraph{Mapping Spoof Features to the Complement.}
As shown in the lower part of Figure~\ref{fig:main2}, the spoof features $h_t^{\mathrm{spf}}\in\mathbb{R}^{D}$ have a different dimension from the Omni features $h_t^{\mathrm{omni}}\in\mathbb{R}^{H}$. Therefore, they cannot be directly represented in the $H$-dimensional Omni space or projected onto $B$. Instead, we use a two-layer MLP to map the spoof features into the coordinate space of the complementary subspace as follows:
\begin{equation}
{h}^{\mathrm{spf}}_{t,B}
= 
\mathrm{MLP}
\left(
{h_t}^{\mathrm{spf}}
\right)
\in
\mathbb{R}^{H-k}.
\end{equation}
For the Omni features, their coordinates in the same subspace are obtained as follows:
\begin{equation}
\ h_{t,B}^{\mathrm{omni}}
=B^\top h_t^{\mathrm{omni}}
\in\mathbb R^{H-k}
\end{equation}
Thus, ${h}^{\mathrm{spf}}_{t,B}$ and ${h}^{\mathrm{omni}}_{t,B}$ are in the same complementary coordinate space and can be fused directly.

\paragraph{Gated Fusion.}
We predict two gates from the Omni and spoof features as follows:
\begin{equation}
{g}^{\mathrm{omni}}
=
\sigma
\left(
\mathrm{MLP}_{\mathrm{omni}}
\left(
[
{h_t}^{\mathrm{omni}};
{h_t}^{\mathrm{spf}}
]
\right)
\right),
\end{equation}
\begin{equation}
{g}^{\mathrm{spf}}
=
\sigma
\left(
\mathrm{MLP}_{\mathrm{spf}}
\left(
[
{h_t}^{\mathrm{omni}};
{h_t}^{\mathrm{spf}}
]
\right)
\right),
\end{equation}
where ${g}^{\mathrm{omni}},{g}^{\mathrm{spf}}\in(0,1)^{H-k}$,
and $\mathrm{MLP}_{\mathrm{omni}}$ and $\mathrm{MLP}_{\mathrm{spf}}$ have independent parameters.
The gate ${g}^{\mathrm{omni}}$ controls how much of the Omni features is retained in the complementary subspace, while ${g}^{\mathrm{spf}}$ controls the amount of spoof information injected into the same subspace. The fused representation is given  as follows:
\begin{equation}
\begin{aligned}
 h_t
= h_t^{\mathrm{omni}}+B\Big[
&( g^{\mathrm{omni}}-\mathbf 1)
\odot h_{t,B}^{\mathrm{omni}}\\
&+ g^{\mathrm{spf}}
\odot h_{t,B}^{\mathrm{spf}}
\Big].
\end{aligned}
\end{equation}
where $\odot$ denotes the Hadamard product, and $+$ denotes element-wise addition.

The final features input into the LLM is ${h}=[{h}_1;\ldots;{h}_T]\in \mathbb{R}^{T\times H}$.
This formulation preserves the components of the original Omni representation in the protected subspace while allowing both representations to be adaptively fused in $\mathrm{span}(B)$. To preserve the original Omni representation at the beginning of training, we initialize the last layer of the spoof MLP to zero and set the bias of the Omni gate such that ${g}^{\mathrm{omni}}\approx\mathbf{1}$. Thus, the initial fused representation satisfies ${h}\approx{h}^{\mathrm{omni}}$, and APSGF starts from the original Omni representation.

\subsection{Reference-Anchored Policy Optimization}
\label{sec:rapo}

For each training example $x$, we sample a single completion $y=(y_1,\ldots,y_L)$ from the current policy $\pi_\theta(\cdot\mid x)$. 

We first remove the spoof tags and use JiWER\footnote{\url{https://jitsi.github.io/jiwer}} to align the reference and predicted transcripts at the word level and compute $\mathrm{WER}{\mathrm{asr}}$. We then assign binary spoof labels to words based on the tags and define $\mathrm{WER}{\mathrm{spoof}}$ as the proportion of alignment steps with mismatched spoof labels. We evaluate each completion using the sentence-level reward  as follows:
\begin{equation}
r(y)
=
1-\frac{
\mathrm{WER}_{\mathrm{asr}}
+
\mathrm{WER}_{\mathrm{spoof}}
}{2},
\end{equation}

We use the post-cross-entropy model as a frozen reference policy
$\pi_{\mathrm{ref}}$. Before policy optimization, we generate a
greedy completion $y_{\mathrm{ref}}$ for each training example and
precompute its reward $r_{\mathrm{ref}}=r(y_{\mathrm{ref}})$.
The reference reward remains fixed throughout training. For each
sampled completion $y$, we define the reference-anchored advantage  as follows:
\begin{equation}
A=r(y)-r_{\mathrm{ref}}.
\end{equation}

Let $L$ denote the number of tokens in the completion $y$. We optimize the completion by the length-normalized sequence log-probability as follows:
\begin{equation}
\log\bar{\pi}_{\theta}(y\mid x)
=
\frac{1}{L}\sum_{t=1}^{L}
\log\pi_{\theta}(y_t\mid x,y_{<t}),
\end{equation}
yielding the policy-gradient loss as follows:
\begin{equation}
L_{\mathrm{RAPO}}
=
-A\log\bar{\pi}_{\theta}(y\mid x).
\end{equation}
A positive advantage reinforces the sampled completion, whereas a negative advantage suppresses it. Moreover we use Kullback--Leibler divergence regularization to constrain policy updates. The final loss is as follows:
\begin{equation}
L
=
L_{\mathrm{RAPO}}
+
D_{\mathrm{KL}}
\left(
\pi_\theta(\cdot\mid x)
\|
\pi_{\mathrm{ref}}(\cdot\mid x)
\right).
\end{equation}
We update $\theta$ by minimizing $L$.

\section{Experiments and Results}
\subsection{Experimental Setup}

\paragraph{Datasets}
We use four English datasets for training: the training set of VCTK\cite{vctk}, PartialEdit\cite{PartialEdit}, AV-Deepfake1M\cite{AV}, and SCDF\cite{Scdf}, with more than 330,000 utterances in total. VCTK provides bona fide speech, while PartialEdit and AV-Deepfake1M provide word-level partial spoofing speech. SCDF provides fully synthetic speech.

We include bona fide VCTK speech to reduce the risk of learning speaker- or content-dependent spoofing cues, as VCTK speakers and read-speech texts have been widely used as source material for synthetic speech generation and spoofing datasets\cite{Datasets}. In PartialEdit and AV-Deepfake1M, an LLM selects specific semantic words for replacement to generate partial spoofing speech. This design may introduce semantic shortcuts that allow the model to associate certain words with spoofing. To reduce this risk, we include fully synthetic speech from SCDF to diversify the relationship between lexical content and spoof labels, with the aim of reducing reliance on word-specific editing patterns.

For evaluation, we use the eval sets of VCTK, PartialEdit, AV-Deepfake1M, SCDF, and LlamaPartialSpoof\cite{llamapartialspoof}. LlamaPartialSpoof is reserved for out-of-domain evaluation. LlamaPartialSpoof contains more than 30,000 utterances, with partial edits generated by six TTS systems. This provides a broader evaluation of generalization to unseen partial-spoofing conditions. Detailed dataset split and statistics are given in Appendix \ref{appendix:datasets}.

\paragraph{Data Processing}

All audio is converted to mono and resampled to 16~kHz. Utterances longer than 40~s are truncated to the first 40~s, and the reference transcripts are truncated to the corresponding leading portion while preserving the spoof labels. We apply RawBoost\cite{rawboost} for waveform augmentation during training.
During ASR-Protected Subspace Gated Fusion, we retain the leading singular vectors whose cumulative explained variance reaches 0.95, which gives $k=175$, to construct the ASR-protected subspace.

\paragraph{Training}

Training consists of two stages: supervised fine-tuning (SFT) and reinforcement learning (RL). We use AdamW with $\beta_1=0.9$, $\beta_2=0.95$, and weight decay 0.01.
The audio encoder is frozen during training, while the language model is updated using DoRA\cite{dora}. DoRA is applied with rank 64, $\alpha=128$, and dropout 0.05. During feature fusion, both gating networks remain trainable and are updated jointly with the language model.
SFT is performed for 5 epochs with token-level cross-entropy loss. The learning rate is $1\times10^{-5}$ with a cosine decay schedule and a warmup ratio of 0.05. 
RL is initialized from the epoch-5 SFT checkpoint with a fresh AdamW optimizer state. The learning rate is $5\times10^{-6}$ with a cosine decay schedule and a warmup ratio of 0.03. Training is performed for 1 epoch.
Temperature is set to $0.5$, top-$p$ is $0.95$, and top-$k$ is $50$. The low temperature and truncated candidate set keep the sample close to greedy decoding while leaving a small amount of stochasticity.

We consider two training settings. The \textbf{two-task setting} jointly performs ASR and word-level spoof detection, where spoofed words are marked with \texttt{<TOF>} and \texttt{<EOF>}. The \textbf{multi-task setting} additionally predicts the utterance-level spoof scope, namely \texttt{none}, \texttt{partial}, or \texttt{all}, before generating the tagged transcript. The detailed training prompts are provided in Appendix \ref{appendix:prompts}.

\paragraph{Evaluation Metrics}
We evaluate the performance of ASR, word-level spoof detection, and utterance-level spoof detection. JiWER\footnote{https://jitsi.github.io/jiwer/}
is used to align words from the reference and the hypothesis based on Levenshtein distance. A word is considered correct when the reference and prediction have the same spoof-tag wrapping, and incorrect otherwise.
For ASR, we report corpus-level Word Error Rate (WER).
For word-level spoof detection, we report macro WordF1, calculated as the average of the F1 scores for the spoof and non-spoof classes.
For utterance-level spoof detection, we report classification accuracy over the three spoof scopes: \texttt{none}, \texttt{partial}, and \texttt{all}. In the multi-task setting, the spoof scope is directly predicted by the model. In the two-task setting, the spoof scope is inferred from the predicted word-level spoof tags. WordF1 is not reported on VCTK because it contains no spoofed words.

\subsection{Experimental Results}
\paragraph{Comparative experiment}

\begin{table*}[t]
\centering
\caption{Comparison of different methods for word-level spoof detection on AV-Deepfake1M (AV), PartialEdit (PE), SCDF (SC), VCTK (VC), and LlamaPartialSpoof (LPS). Each entry is \textbf{WER$\downarrow$ / WordF1$\uparrow$}. The Avg. column is the five-set mean of WER and the four-set mean of WordF1.}
\begin{tabular}{l|l l l l l |l}
\toprule
\textbf{Models}               & \textbf{AV} & \textbf{PE} & \textbf{SC} & \textbf{VC} & \textbf{LPS} & \textbf{Avg.} \\ \midrule
SLM-A & 0.116/0.947 &0.629/0.541 &------/------ &------/------ &------/------ &0.373/0.744 \\ 
HoliAS                  & ------/0.210       & ------/0.827            & ------/0.222     & ------/------       & ------/0.254                  & ------/0.378    \\ 
W\_TTS*\tnote{*}                   & 0.159/0.131     & 0.050/0.287         & 0.214/0.884   & 0.032/------    & \textbf{0.067}/\textbf{0.886}               & 0.104/0.547  \\ 
W\_VT                    & 0.141/0.223     & 0.043/0.321         & 0.218/0.746   & 0.023/------    & 0.068/0.884               & 0.099/0.543 \\ 
\textbf{Ours\_T} & 0.081/0.969    & 0.032/0.853         & 0.114/0.934   & 0.010/------   & 0.110/0.838                & 0.069/0.898 \\
\textbf{Ours\_M}     & \textbf{0.065}/\textbf{0.975}    & \textbf{0.024/0.907}         & \textbf{0.089/0.964}  & \textbf{0.004}/------   & 0.086/0.759               & \textbf{0.054/0.901} \\ 
\bottomrule
\end{tabular}
\begin{tablenotes}
\footnotesize
\item[*]*Finetuned on LlamaPartialSpoof.
\end{tablenotes}
\label{tab:res_word_comp}
\end{table*}

\begin{table}[t]
\centering
\caption{Comparison of different methods for utterance-level spoof scope classification on AV-Deepfake1M (AV), PartialEdit (PE), SCDF (SC), VCTK (VC), and LlamaPartialSpoof (LPS). Each entry denotes the utterance-level classification accuracy (\textbf{U-ACC$\uparrow$ }), and the Avg. reports the average across the five datasets.}
\setlength{\tabcolsep}{2.5pt}
\begin{tabular}{@{}l|ccccc|c@{}}
\toprule
\textbf{Models} & \textbf{AV} & \textbf{PE} & \textbf{SC} &
\textbf{VC} & \textbf{LPS} & \textbf{Avg.} \\
\midrule

HoliAS      & 0.666           & \textbf{0.981 }               & 0.866         & 0.866         & 0.615            & 0.799        \\
W\_TTS        & 0.859           & 0.269                & 0.688         & 0.004       & \textbf{0.947}            & 0.553        \\
W\_VT        & 0.888           & 0.568                & 0.403         & 0.561         & 0.946            & 0.673        \\
\textbf{Ours\_T} & 0.960           & 0.811                & 0.904         & 0.993         & 0.454            & 0.824        \\
\textbf{Ours\_M} & \textbf{1.000}            & 0.974                & \textbf{0.983}         & \textbf{1.000 }         & 0.805            & \textbf{0.952}        \\  
\bottomrule
\end{tabular}
\label{tab:res_utt_comp}
\end{table}

We compare two variants of our model with three open-source models. \textbf{SLM-A} is from \citet{zhang2026can}, which uses a WavLM-based speech encoder\cite{wavlm} and TinyLlama\cite{llama} to jointly transcribe speech and localize fake words, trained separately on the training splits of PartialEdit and AV-Deepfake1M. \textbf{HoliAS} is HoliAntiSpoof\cite{xu2026holiantispoof}. It predicts whether an utterance is real or fake, the spoofing method, the fake region, and a natural-language description of the semantic influence, rather than a transcript, so WER is left blank. But we can capture the detected spoof words in its semantic influence sentence to compute the WordF1. Following the training setup of HoliAntiSpoof, among the five evaluation sets, PartialEdit and VCTK are in-domain. AV-Deepfake1M, SCDF, and LlamaPartialSpoof are out-of-domain. \textbf{W\_TTS} \cite{tran26_interspeech} is Whisper-large \cite{whisper} fine-tuned for next-token word detection on LlamaPartialSpoof. LlamaPartialSpoof is therefore in-domain for W\_TTS, and the other four sets are out-of-domain. \textbf{W\_VT} is Whisper-large fine-tuned on vocoded Multilingual LibriSpeech mixed with TTS spoofs. None of the five evaluation sets is in our training mixture. \textbf{Ours\_T} is the two-task model and \textbf{Ours\_M} is the multi-task model. Both are trained on AV-Deepfake1M, PartialEdit, SCDF, and VCTK, so LlamaPartialSpoof is out-of-domain.

\begin{table*}[t]
\centering
\caption{Ablation of the APSGF and RAPO under the multi-task settings on AV-Deepfake1M (AV), PartialEdit (PE), SCDF (SC), VCTK (VC), and LlamaPartialSpoof (LPS). Each entry is\textbf{ WER$\downarrow$ / WordF1$\uparrow$}. The Avg. column is the five-set mean of WER and the four-set mean of WordF1.}
\begin{tabular}{c c|l l l l l  |l}
\toprule
 \textbf{APSGF} & \textbf{RAPO}
& \textbf{AV} & \textbf{PE} & \textbf{SC}
& \textbf{VC} & \textbf{LPS} & \textbf{Avg.} \\

\midrule
    \xmark &\xmark                   & 0.086/0.967 & 0.026/0.873 & 0.136/0.938 & 0.007/---   & 0.088/0.720  & 0.069/0.874          \\
\cmark &\xmark                                 & 0.061/0.977  & 0.033/0.907 & 0.096/0.961 & 0.006/---   & 0.093/0.719  & 0.058/0.891          \\
      \xmark &\cmark                                   & 0.072/0.976  & 0.026/0.905 & 0.093/0.962 & 0.007/---   & 0.086/0.760  & 0.057/0.901          \\
\cmark &\cmark                                                 & 0.065/0.975  & 0.024/0.907 & 0.089/0.964 & 0.004/---    & 0.086/0.759  & \textbf{0.054/0.901} \\

\bottomrule
\end{tabular}

\label{tab:ablation}

\end{table*}

\begin{table*}[!t]
\caption{Comparison of APSGF and RAPO with alternative optimization methods under the multi-task setting on AV-Deepfake1M (AV), PartialEdit (PE), SCDF (SC), VCTK (VC), and LlamaPartialSpoof (LPS). Each entry is \textbf{WER$\downarrow$ / WordF1$\uparrow$}. The Avg. column reports the five-dataset mean of WER and the four-dataset mean of WordF1.}
\begin{tabular}{l|l|lllll|l}
\toprule
\textbf{\begin{tabular}[c]{@{}l@{}}Exp.\\ No.\end{tabular}} & 

\textbf{Method} & \textbf{AV}  & \textbf{PE} & \textbf{SC} & \textbf{VC} & \textbf{LPS} & \textbf{Avg.}         \\
\midrule
0                                                           & Base            & 0.086/0.967 & 0.026/0.873 & 0.136/0.938 & 0.007/---   & 0.088/0.720  & 0.069/0.874          \\
1                                                           & Aux. Enc.       & 0.077/0.972  & 0.025/0.873 & 0.137/0.951 & 0.006/---   & 0.081/0.726  & 0.065/0.880          \\
2                                                           & APSGF           & 0.061/0.977  & 0.033/0.907 & 0.096/0.961 & 0.006/---   & 0.093/0.719  & \textbf{0.058/0.891} \\
\midrule
3                                                           & +GRPO            & 0.084/0.969  & 0.025/0.901 & 0.139/0.901 & 0.008/---   & 0.086/0.717  & 0.068/0.872          \\
4                                                           & +RAPO            & 0.065/0.975  & 0.024/0.907 & 0.089/0.964 & 0.004/---    & 0.086/0.759  & \textbf{0.054/0.901} \\
\bottomrule
\end{tabular}
\label{tab:abl2}
\end{table*}

Table~\ref{tab:res_word_comp} compares different methods for word-level spoof detection. W\_TTS achieves 0.886 WordF1 on the in-domain LlamaPartialSpoof set but drops to 0.131--0.287 on the OOD datasets. Similarly, HoliAS reaches 0.827 WordF1 on the in-domain PartialEdit set but drops to 0.210--0.254 on the OOD datasets. 
Our models achieve strong average performance across the five evaluation sets. However, multi-task training does not consistently improve out-of-domain word-level detection; it decreases from 0.838 for the two-task model to 0.759 for the multi-task model. One possible explanation is that scope prediction errors affect subsequent word tags, although this requires further analysis.
On VCTK, Ours\_M achieves a very low WER of 0.004. This low WER should be interpreted with caution given the repeated text in VCTK and its overlap with existing spoofing corpora.

Table~\ref{tab:res_utt_comp} compares different methods for utterance-level spoof scope classification. Ours\_M achieves the best average U-ACC of 0.952, with near-perfect accuracy on the in-domain datasets and 0.805 on the OOD LlamaPartialSpoof. In contrast, Ours\_T averages 0.824 and drops to 0.454 on LlamaPartialSpoof, showing the benefit of directly predicting the spoof scope in the multi-task setting. Among the baselines, HoliAS achieves 0.799 on average, while W\_VT and W\_TTS reach only 0.673 and 0.553. Notably, HoliAS obtains 0.866 U-ACC on SCDF despite a WordF1 of only 0.222, indicating that correct utterance-level scope prediction does not guarantee accurate word-level localization. On PartialEdit, W\_TTS achieves a low WER of 0.050 but a WordF1 of only 0.287, illustrating
that accurate transcription alone does not ensure reliable spoof localization.

\paragraph{Ablation Study}

Table~\ref{tab:ablation} presents the ablation results of APSGF and RAPO. Compared with the baseline, APSGF reduces the average WER from 0.069 to 0.058 and improves the average WordF1 from 0.874 to 0.891. This shows that APSGF integrates spoof features while preserving ASR performance. RAPO alone further improves the average WordF1 to 0.901. Combining both modules achieves the lowest average WER of 0.054 while maintaining a WordF1 of 0.901. These results show APSGF and RAPO improve transcription accuracy while maintaining strong word-level spoof detection performance.

Table~\ref{tab:abl2} compares the auxiliary encoder, APSGF, and different policy optimization methods. Exp. No. 0 is the baseline without an auxiliary encoder, APSGF, or RAPO. Exp. No. 1 adds an auxiliary encoder with simple gated fusion, using the same number of gates as APSGF but without ASR protection. It reduces the average WER from 0.069 to 0.065, showing the value of additional acoustic features. Exp. No. 2 uses APSGF and further reduces the average WER to 0.058 while improving WordF1 to 0.891. This comparison shows that APSGF's gains come not only from the auxiliary feature extractor but also from its ASR-protected subspace fusion mechanism. Exp. No. 3 and 4 apply GRPO and RAPO, respectively, on top of APSGF. Compared with GRPO, RAPO reduces the average WER from 0.068 to 0.054 and improves WordF1 from 0.872 to 0.901, while GRPO brings no clear overall improvement. These results suggest that RAPO is more effective than GRPO for further optimizing an SFT-initialized policy. Details of fusion of the auxiliary encoder, GRPO, and model sizes are provided in Appendices ~\ref{gate}, ~\ref{grpo}, and ~\ref{Parameter}, respectively.

\section{Conclusion and Discussion}

We explored tagged ASR for joint ASR and word-level speech anti-spoofing. APSGF incorporates spoof cues while preserving ASR representations, and RAPO provides stable policy optimization anchored to the post-SFT model. 
Experiments on five datasets demonstrate improvements in average ASR and word-level spoof detection performance. However, the results on LlamaPartialSpoof indicate that further improvements in out-of-domain word-level detection remain challenging.
Ablations show complementary gains from the auxiliary encoder, APSGF, and RAPO in spoof representation, localization, and transcription.

The framework can extend to other acoustic-semantic tasks, such as speech understanding and fine-grained audio analysis. RAPO may also benefit tasks where the model changes little after SFT, using the reference output as a stable optimization anchor. Together, these methods offer a way to add acoustic objectives to well-trained audio-language models while preserving their existing capabilities.

\clearpage

\section*{Limitations}

Our study has several limitations. First, we evaluate the proposed framework only on word-level speech anti-spoofing and do not validate its applicability to other acoustic-semantic tasks. Second, experiments are conducted with a single audio-language model, so the generality of APSGF and RAPO across different model architectures remains unclear. Third, the current tagged ASR formulation mainly identifies spoofed words and cannot detect deletion-based edits. Finally, the experiments use limited hyperparameter tuning, and further optimization may lead to additional gains.
\section{Ethical Considerations}
The datasets used in this work are publicly available and are used in accordance with their licenses and terms. Large Language Models (LLMs) were used solely for manuscript polishing to improve clarity and readability. The LLMs were not used for ideation, methodology, experimental design, data analysis, or result interpretation. All scientific content was produced and verified by the authors.

\section*{Acknowledgments}
We thank Hoan My TRAN, Jin Li, Xuechen Liu, and Xin Wang for helpful discussions.

\bibliography{custom}

@ARTICLE{partialspoof,
  author={Zhang, Lin and Wang, Xin and Cooper, Erica and Evans, Nicholas and Yamagishi, Junichi},
  journal={IEEE/ACM Transactions on Audio, Speech, and Language Processing}, 
  title={The PartialSpoof Database and Countermeasures for the Detection of Short Fake Speech Segments Embedded in an Utterance}, 
  year={2023},
  volume={31},
  number={},
  pages={813-825},
  doi={10.1109/TASLP.2022.3233236}}

@article{xu2026holiantispoof,
  title={Holiantispoof: Audio llm for holistic speech anti-spoofing},
  author={Xu, Xuenan and Ren, Yiming and Liu, Liwei and Wu, Wen and Li, Baoxiang and Lu, Chaochao and Wang, Shuai and Zhang, Chao},
  journal={arXiv preprint arXiv:2602.04535},
  year={2026},
  doi={10.48550/arXiv.2602.04535}
}

@article{xue2026unifying,
  title={Unifying Speech Editing Detection and Content Localization via Prior-Enhanced Audio LLMs},
  author={Xue, Jun and Chai, Yi and Ren, Yanzhen and He, Jinshen and Tang, Zhiqiang and Yi, Zhuolin and Huang, Yihuan and Xie, Yuankun and Chen, Yujie},
  journal={arXiv preprint arXiv:2601.21463},
  year={2026},
  doi={10.48550/arXiv.2601.21463}
}

@inproceedings{tran26_interspeech,
  title     = {{Deepfake Word Detection by Next-token Prediction using Fine-tuned Whisper}},
  author    = {Hoan My Tran and Xin Wang and Wanying Ge and Xuechen Liu and Junichi Yamagishi},
  year      = {2026},
  booktitle = {{Interspeech}},
  pages     = {1042--1046},
  doi       = {10.21437/Interspeech.2026-628},
  issn      = {2958-1796},
}

@article{zhang2026can,
  title={Can LLMs Help Localize Fake Words in Partially Fake Speech?},
  author={Zhang, Lin and Thebaud, Thomas and Cai, Zexin and Khudanpur, Sanjeev and Povey, Daniel and Garc{\'\i}a-Perera, Leibny Paola and Wiesner, Matthew and Andrews, Nicholas},
  journal={arXiv preprint arXiv:2603.11205},
  year={2026},
  doi={10.48550/arXiv.2603.11205}
}

@inproceedings{tang2024salmonn,
  title={Salmonn: Towards generic hearing abilities for large language models},
  author={Tang, Changli and Yu, Wenyi and Sun, Guangzhi and Chen, Xianzhao and Tan, Tian and Li, Wei and Lu, Lu and Ma, Zejun and Zhang, Chao},
  booktitle={International Conference on Learning Representations},
  volume={2024},
  pages={16607--16629},
  year={2024},
  url={https://proceedings.iclr.cc/paper_files/paper/2024/hash/476ab8f369e489c04187ba84f68cfa68-Abstract-Conference.html}
}

@inproceedings{2024wavllm,
    title = "{W}av{LLM}: Towards Robust and Adaptive Speech Large Language Model",
    author = "Hu, Shujie  and
      Zhou, Long  and
      Liu, Shujie  and
      Chen, Sanyuan  and
      Meng, Lingwei  and
      Hao, Hongkun  and
      Pan, Jing  and
      Liu, Xunying  and
      Li, Jinyu  and
      Sivasankaran, Sunit  and
      Liu, Linquan  and
      Wei, Furu",
    editor = "Al-Onaizan, Yaser  and
      Bansal, Mohit  and
      Chen, Yun-Nung",
    booktitle = "Findings of Empirical Methods in Natural Language Processing",
    month = nov,
    year = "2024",
    url = "https://aclanthology.org/2024.findings-emnlp.263/",
    doi = "10.18653/v1/2024.findings-emnlp.263",
    pages = "4552--4572",
}

@inproceedings{2025enhancing,
    title = "Enhancing Speech Large Language Models with Prompt-Aware Mixture of Audio Encoders",
    author = "Shan, Weiqiao  and
      Li, Yuang  and
      Zhang, Yuhao  and
      Luo, Yingfeng  and
      Xu, Chen  and
      Zhao, Xiaofeng  and
      Meng, Long  and
      Lu, Yunfei  and
      Zhang, Min  and
      Yang, Hao  and
      Xiao, Tong  and
      Zhu, JingBo",
    editor = "Christodoulopoulos, Christos  and
      Chakraborty, Tanmoy  and
      Rose, Carolyn  and
      Peng, Violet",
    booktitle = "Empirical Methods in Natural Language Processing",
    month = nov,
    year = "2025",
    url = "https://aclanthology.org/2025.emnlp-main.974/",
    doi = "10.18653/v1/2025.emnlp-main.974",
    pages = "19305--19320",
    ISBN = "979-8-89176-332-6",
   
}

@inproceedings{zhang2025mowe,
  title={Mowe-audio: Multitask audiollms with mixture of weak encoders},
  author={Zhang, Wenyu and Sun, Shuo and Wang, Bin and Zou, Xunlong and Liu, Zhuohan and He, Yingxu and Lin, Geyu and Chen, Nancy F and Aw, Ai Ti},
  booktitle={IEEE International Conference on Acoustics, Speech and Signal Processing},
  pages={1--5},
  year={2025},
  doi={10.1109/ICASSP49660.2025.10888128}
}

@inproceedings{gao2024speech,
  title={Speech emotion recognition with multi-level acoustic and semantic information extraction and interaction},
  author={Gao, Yuan and Shi, Hao and Chu, Chenhui and Kawahara, Tatsuya},
  booktitle={Interspeech},
  pages={1060--1064},
  year={2024},
  doi       = {10.21437/Interspeech.2024-2385},
}

@inproceedings{wang25u_interspeech,
  title     = {{Continual Speech Learning with Fused Speech Features}},
  author    = {Guitao Wang and Jinming Zhao and Hao Yang and Guilin Qi and Tongtong Wu and Gholamreza Haffari},
  year      = {2025},
  booktitle = {{Interspeech}},
  pages     = {1793--1797},
  doi       = {10.21437/Interspeech.2025-1823},
  issn      = {2958-1796},
}

@article{rafailov2023direct,
  title={Direct preference optimization: Your language model is secretly a reward model},
  author={Rafailov, Rafael and Sharma, Archit and Mitchell, Eric and Manning, Christopher D and Ermon, Stefano and Finn, Chelsea},
  journal={Advances in neural information processing systems},
  volume={36},
  pages={53728--53741},
  year={2023},
  doi={10.52202/075280-2338}
}

@article{meng2024simpo,
  title={Simpo: Simple preference optimization with a reference-free reward},
  author={Meng, Yu and Xia, Mengzhou and Chen, Danqi},
  journal={Advances in Neural Information Processing Systems},
  volume={37},
  pages={124198--124235},
  year={2024},
  doi={10.52202/079017-3946}
}

@article{shao2024deepseekmath,
  title={Deepseekmath: Pushing the limits of mathematical reasoning in open language models},
  author={Shao, Zhihong and Wang, Peiyi and Zhu, Qihao and Xu, Runxin and Song, Junxiao and Bi, Xiao and Zhang, Haowei and Zhang, Mingchuan and Li, YK and others},
  journal={arXiv preprint arXiv:2402.03300},
  year={2024},
  doi={10.48550/arXiv.2402.03300}
}

@article{yu2026dapo,
  title={Dapo: An open-source llm reinforcement learning system at scale},
  author={Yu, Qiying and Zhang, Zheng and Zhu, Ruofei and Yuan, Yufeng and Zuo, Xiaochen and Yue, Yu and Dai, Weinan and Fan, Tiantian and Liu, Gaohong and Liu, Lingjun and others},
  journal={Advances in Neural Information Processing Systems},
  volume={38},
  pages={113222--113244},
  year={2026},
  doi={10.52202/085713-3775}
}

@article{ppo,
  title={Proximal policy optimization algorithms},
  author={Schulman, John and Wolski, Filip and Dhariwal, Prafulla and Radford, Alec and Klimov, Oleg},
  journal={arXiv preprint arXiv:1707.06347},
  year={2017},
  doi={10.48550/arXiv.2602.04535}
}

@misc{qwen25omni,
      title={Qwen2.5-Omni Technical Report}, 
      author={Jin Xu and Zhifang Guo and Jinzheng He and Hangrui Hu and Ting He and Shuai Bai and Keqin Chen and Jialin Wang and Yang Fan and Kai Dang and Bin Zhang and Xiong Wang and Yunfei Chu and Junyang Lin},
      year={2025},
      eprint={2503.20215},
      archivePrefix={arXiv},
      primaryClass={cs.CL},
      url={https://arxiv.org/abs/2503.20215}, 
}

@inproceedings{opd,
  title={On-policy distillation of language models: Learning from self-generated mistakes},
  author={Agarwal, Rishabh and Vieillard, Nino and Zhou, Yongchao and Stanczyk, Piotr and Ramos Garea, Sabela and Geist, Matthieu and Bachem, Olivier},
  booktitle={International Conference on Learning Representations},
  volume={2024},
  pages={21246--21263},
  year={2024},
  url={https://proceedings.iclr.cc/paper_files/paper/2024/file/5be69a584901a26c521c2b51e40a4c20-Paper-Conference.pdf}
}

@InProceedings{remax,
  title = 	 {{R}e{M}ax: A Simple, Effective, and Efficient Reinforcement Learning Method for Aligning Large Language Models},
  author =       {Li, Ziniu and Xu, Tian and Zhang, Yushun and Lin, Zhihang and Yu, Yang and Sun, Ruoyu and Luo, Zhi-Quan},
  booktitle = 	 {International Conference on Machine Learning},
  pages = 	 {29128--29163},
  year = 	 {2024},
  editor = 	 {Salakhutdinov, Ruslan and Kolter, Zico and Heller, Katherine and Weller, Adrian and Oliver, Nuria and Scarlett, Jonathan and Berkenkamp, Felix},
  volume = 	 {235},
  month = 	 {21--27 Jul},
  url = 	 {https://proceedings.mlr.press/v235/li24cd.html},
}

@article{guo2025deepseek,
  title={DeepSeek-R1 incentivizes reasoning in LLMs through reinforcement learning},
  author={Guo, Daya and Yang, Dejian and Zhang, Haowei and Song, Junxiao and Wang, Peiyi and Zhu, Qihao and Xu, Runxin and Zhang, Ruoyu and Ma, Shirong and Bi, Xiao and others},
  journal={Nature},
  volume={645},
  number={8081},
  pages={633--638},
  year={2025},
  doi={10.1038/s41586-025-09422-z}
}

@inproceedings{shenfeld2026rl,
  title={Rl's razor: Why online reinforcement learning forgets less},
  author={Shenfeld, Idan and Pari, Jyothish and Agrawal, Pulkit},
  booktitle={International Conference on Learning Representations},
  volume={2026},
  pages={59839--59864},
  year={2026},
  url={https://proceedings.iclr.cc/paper_files/paper/2026/file/618c95f4557c15b253fb0e6f548ea0c0-Paper-Conference.pdf}
}

@misc{vctk,
  author = {Yamagishi, Junichi and Veaux, Christophe and MacDonald, Kirsten},
  title = {{CSTR} {VCTK} Corpus: English Multi-speaker Corpus for {CSTR} Voice Cloning Toolkit},
  year = {2019},
  doi = {10.7488/ds/2645}
}

@inproceedings{Scdf,
  title={Scdf: A speaker characteristics deepfake speech dataset for bias analysis},
  author={Stan{\v{e}}k, Vojt{\v{e}}ch and Srna, Karel and Firc, Anton and Malinka, Kamil},
  booktitle={International Conference of the Biometrics Special Interest Group},
  pages={1--6},
  year={2025},
  doi={10.1109/BIOSIG65492.2025.11358014}
}

@inproceedings{PartialEdit,
  title     = {{PartialEdit: Identifying Partial Deepfakes in the Era of Neural Speech Editing  }},
  author    = {You Zhang and Baotong Tian and Lin Zhang and Zhiyao Duan},
  year      = {2025},
  booktitle = {{Interspeech}},
  pages     = {5353--5357},
  doi       = {10.21437/Interspeech.2025-942},
  issn      = {2958-1796},
}

@inproceedings{llamapartialspoof,
  title={Llamapartialspoof: An llm-driven fake speech dataset simulating disinformation generation},
  author={Luong, Hieu-Thi and Li, Haoyang and Zhang, Lin and Lee, Kong Aik and Chng, Eng Siong},
  booktitle={IEEE International Conference on Acoustics, Speech and Signal Processing},
  pages={1--5},
  year={2025},
  doi={10.1109/ICASSP49660.2025.10888070}
}

@inproceedings{AV,
  title={AV-Deepfake1M: A large-scale LLM-driven audio-visual deepfake dataset},
  author={Cai, Zhixi and Ghosh, Shreya and Adatia, Aman Pankaj and Hayat, Munawar and Dhall, Abhinav and Gedeon, Tom and Stefanov, Kalin},
  booktitle={ACM International Conference on Multimedia},
  pages={7414--7423},
  year={2024},
  doi={10.1145/3664647.368079}
}

@inproceedings{Datasets,
  title     = {{Ethical and Technical Limits of Deepfake Speech Datasets}},
  author    = {Vojtěch Staněk and Eva Trnovská and Kamil Malinka and Anton Firc},
  year      = {2026},
  booktitle = {{Interspeech}},
  pages     = {2340--2345},
  doi       = {10.21437/Interspeech.2026-124},
  issn      = {2958-1796},
}

@inproceedings{rawboost,
  title={Rawboost: A raw data boosting and augmentation method applied to automatic speaker verification anti-spoofing},
  author={Tak, Hemlata and Kamble, Madhu and Patino, Jose and Todisco, Massimiliano and Evans, Nicholas},
  booktitle={IEEE International Conference on Acoustics, Speech and Signal Processing},
  pages={6382--6386},
  year={2022},
  doi={10.1109/ICASSP43922.2022.9746213}
}

@inproceedings{dora,
author = {Liu, Shih-Yang and Wang, Chien-Yi and Yin, Hongxu and Molchanov, Pavlo and Wang, Yu-Chiang Frank and Cheng, Kwang-Ting and Chen, Min-Hung},
title = {DoRA: weight-decomposed low-rank adaptation},
year = {2024},
booktitle = {International Conference on Machine Learning},
articleno = {1299},
numpages = {22},
volume = {235},
pages = {32100--32121},
url={https://proceedings.mlr.press/v235/liu24bn.html}
}

@inproceedings{whisper,
  title={Robust speech recognition via large-scale weak supervision},
  author={Radford, Alec and Kim, Jong Wook and Xu, Tao and Brockman, Greg and McLeavey, Christine and Sutskever, Ilya},
  booktitle={International Conference on Machine Learning},
  pages={28492--28518},
  year={2023},
  url={https://proceedings.mlr.press/v202/radford23a.html}
}

@article{wavlm,
  title={Wavlm: Large-scale self-supervised pre-training for full stack speech processing},
  author={Chen, Sanyuan and Wang, Chengyi and Chen, Zhengyang and Wu, Yu and Liu, Shujie and Chen, Zhuo and Li, Jinyu and Kanda, Naoyuki and Yoshioka, Takuya and Xiao, Xiong and others},
  journal={IEEE Journal of Selected Topics in Signal Processing},
  volume={16},
  number={6},
  pages={1505--1518},
  year={2022},
  doi={10.1109/JSTSP.2022.3188113}
}

@misc{llama,
      title={TinyLlama: An Open-Source Small Language Model}, 
      author={Peiyuan Zhang and Guangtao Zeng and Tianduo Wang and Wei Lu},
      year={2024},
      eprint={2401.02385},
      archivePrefix={arXiv},
      primaryClass={cs.CL},
      doi={10.48550/arXiv.2401.02385}
}

@article{cai2024integrating,
  title={Integrating frame-level boundary detection and deepfake detection for locating manipulated regions in partially spoofed audio forgery attacks},
  author={Cai, Zexin and Li, Ming},
  journal={Computer Speech \& Language},
  volume={85},
  pages={101597},
  year={2024},
  url = {https://doi.org/10.1016/j.csl.2023.101597},
}

@InProceedings{BEATs,
  title = 	 {{BEAT}s: Audio Pre-Training with Acoustic Tokenizers},
  author =       {Chen, Sanyuan and Wu, Yu and Wang, Chengyi and Liu, Shujie and Tompkins, Daniel and Chen, Zhuo and Che, Wanxiang and Yu, Xiangzhan and Wei, Furu},
  booktitle = 	 {Proceedings of the 40th International Conference on Machine Learning},
  pages = 	 {5178--5193},
  year = 	 {2023},
  volume = 	 {202},
  series = 	 {Proceedings of Machine Learning Research},
  month = 	 {23--29 Jul},
  url = 	 {https://proceedings.mlr.press/v202/chen23ag.html},
}

@inproceedings{gpm,
title={Gradient Projection Memory for Continual Learning},
author={Gobinda Saha and Isha Garg and Kaushik Roy},
booktitle={International Conference on Learning Representations},
year={2021},
URL={https://openreview.net/forum?id=3AOj0RCNC2}
}

@Article{SOP,
AUTHOR = {Cheng, Su and Liu, Guoyang},
TITLE = {SOP: Selective Orthogonal Projection for Composed Image Retrieval},
JOURNAL = {Sensors},
VOLUME = {26},
YEAR = {2026},
NUMBER = {5},
ARTICLE-NUMBER = {1621},
URL = {https://www.mdpi.com/1424-8220/26/5/1621},
PubMedID = {41829582},
ISSN = {1424-8220},
DOI = {10.3390/s26051621}
}

@article{sd,
  title={A survey on speech deepfake detection},
  author={Li, Menglu and Ahmadiadli, Yasaman and Zhang, Xiao-Ping},
  journal={ACM Computing Surveys},
  volume={57},
  number={7},
  pages={1--38},
  year={2025},
  URL={https://dl.acm.org/doi/10.1145/3714458}
}

@article{asv,
  title={ASVspoof 2021: accelerating progress in spoofed and deepfake speech detection},
  author={Yamagishi, Junichi and Wang, Xin and Todisco, Massimiliano and Sahidullah, Md and Patino, Jose and Nautsch, Andreas and Liu, Xuechen and Lee, Kong Aik and Kinnunen, Tomi and Evans, Nicholas and others},
  journal={arXiv preprint arXiv:2109.00537},
  year={2021},
  doi={10.48550/arXiv.2109.00537}
}

\appendix

\begin{table*}[]
    \begin{center}

\captionof{table}{Statistics of the training and evaluation datasets. ``Partial'' and ``Full'' denote partially and fully spoofed utterances, respectively.}
\label{tab:datasets}

\resizebox{\textwidth}{!}{%
\begin{tabular}{l l r r r r r l}
\toprule
\textbf{Dataset} & \textbf{Sub Set} & \textbf{Split} & \textbf{\# Utt.} &
\textbf{Bona fide} & \textbf{Partial} & \textbf{Full} & \textbf{Role} \\
\midrule
VCTK & Full & Train & 76,849 & 76,849 & -- & -- & Training \\
PartialEdit & E1, E2 &Train & 16,524 & -- & 16,524 & -- & Training \\
AV-Deepfake1M & Fake audio & Train & 199,895 & -- & 199,895 & -- & Training \\
SCDF & EN & 4/5 & 37,960 & -- & -- & 37,960 & Training \\
\midrule
VCTK & Full & Val. & 4,028 & 4,028 & -- & -- & Evaluation \\
PartialEdit & E1, E2 &Eval & 55,976 & -- & 55,976 & -- & Evaluation \\
AV-Deepfake1M & Fake audio & Val & 27,837 & -- & 27,837 & -- & Evaluation \\
SCDF & EN & 1/5 & 9,490 & -- & -- & 9,490 & Evaluation \\
LlamaPartialSpoof & R01TTS.0.b &Dev & 32,194 & -- & 32,194 & -- & Evaluation-OOD \\
\bottomrule
\end{tabular}%
}
\end{center}
\end{table*}


\section{Dataset Details}
\label{appendix:datasets}

Table~\ref{tab:datasets} summarizes the datasets used for training and evaluation. We use only English data with available transcriptions; therefore, only a subset of some datasets is included. For VCTK, we use the data and splits provided in the public Hugging Face repository\footnote{\url{https://huggingface.co/datasets/badayvedat/VCTK}}. For PartialEdit, we use only the publicly available E1 and E2 subsets. For AV-Deepfake1M, we select only audio samples containing speech deepfakes. Since SCDF does not provide official training and test splits, we randomly assign four-fifths of the data to training and the remaining one-fifth to testing. The training set comprises VCTK, PartialEdit, AV-Deepfake1M, and SCDF, covering bona fide, partially spoofed, and fully spoofed utterances. For evaluation, we use the corresponding held-out splits, while the R01TTS.0.b subset of LlamaPartialSpoof serves as the out-of-domain (OOD) test set.

\section{Training Prompts}
\label{appendix:prompts}

\subsection{Two-Task Setting}

\textbf{Prompt.}
\[
\fbox{\parbox{0.93\linewidth}{
Transcribe the audio and identify which words are spoofed. Wrap spoofed words with \texttt{<TOF>} and \texttt{<EOF>}. Only wrap spoofed words. Do not explain. If none, output plain transcription.
}}
\]

\textbf{Reference answers.}
\[
\fbox{\parbox{0.93\linewidth}{
\textbf{none:} Please call Stella.

\textbf{partial:} Please \texttt{<TOF>}ignore\texttt{<EOF>} Stella.
\parbox{0.93\linewidth}{
\raggedright
\textbf{all:} \texttt{<TOF>}Please\texttt{<EOF>} \texttt{<TOF>}ignore\texttt{<EOF>} \texttt{<TOF>}Stella\texttt{<EOF>}.
}
}}
\]

\subsection{Multi-Task Setting}

\textbf{Prompt.}
\[
\fbox{\parbox{0.93\linewidth}{
Classify the audio's spoof level. Then transcribe the audio and identify which words are spoofed. Wrap spoofed words with \texttt{<TOF>} and \texttt{<EOF>}.
}}
\]

\textbf{Reference answers.}
\[
\fbox{\parbox{0.93\linewidth}{
\textbf{none:}\\
Spoof level: none\\
Transcript: Please call Stella.

\textbf{partial:}\\
Spoof level: partial\\
Transcript: Please \texttt{<TOF>}ignore\texttt{<EOF>} Stella.

\textbf{all:}\\
Spoof level: all\\
\parbox{0.93\linewidth}{
\raggedright
Transcript: \texttt{<TOF>}Please\texttt{<EOF>} \texttt{<TOF>}ignore \\ \texttt{<EOF>} \texttt{<TOF>}Stella\texttt{<EOF>}.
}}
}
\]

\begin{figure}[t]
    \centering
    \includegraphics[width=\linewidth]{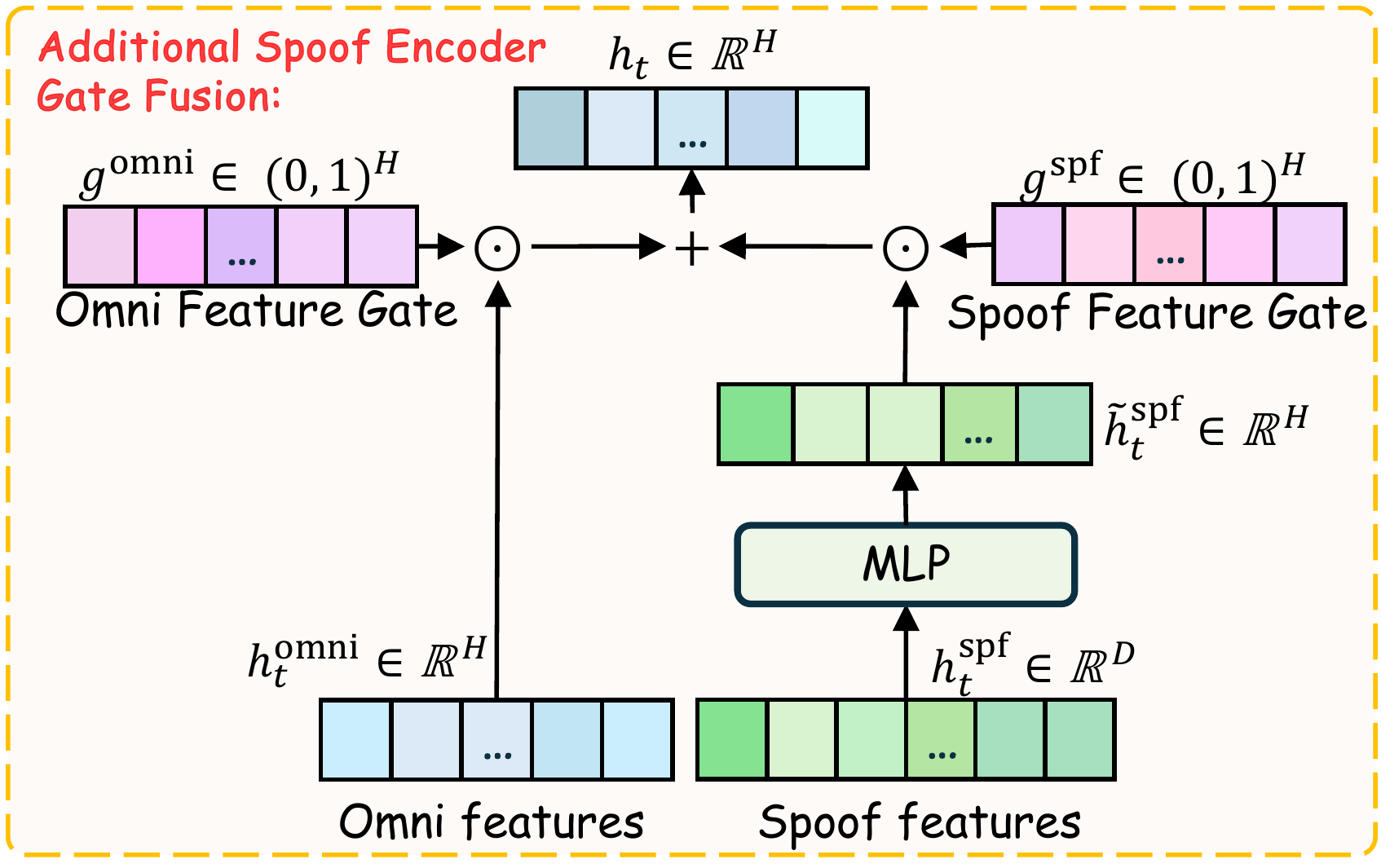}
    \caption{Overview of Additional Spoof Encoder Gate Fusion Details. Two independent gates control the contributions of Omni and spoof features. The gated features are then added across the full Omni feature space to obtain the fused representation.}
    \label{fig:main3}
\end{figure}

\section{Auxiliary Spoof Encoder Gate Fusion Details}
\label{gate}
As a baseline for comparison, we replace APSGF with a simple gated fusion mechanism. As shown in ~\ref{fig:main3}, the aligned spoof features re denoted by $h^{\mathrm{spf}}\in\mathbb{R}^{T\times D}$ and the Omni features by $h^{\mathrm{omni}}\in\mathbb{R}^{T\times H}$. At each temporal position  $t$, we first map the spoof feature $h^{\mathrm{spf}}_t\in\mathbb{R}^{D}$ to the Omni feature dimension using an MLP:
\begin{equation}
    \tilde{h}^{\mathrm{spf}}_t
    = \operatorname{MLP}
    (h^{\mathrm{spf}}_t)
    \in\mathbb{R}^{H}.
\end{equation}

We then predict two independent gates from the
concatenated Omni and spoof features:
\begin{equation}
    g^{\mathrm{omni}}_t
    = \sigma\left(
    \operatorname{MLP}_{\mathrm{omni}}
    ([h^{\mathrm{omni}}_t;h^{\mathrm{spf}}_t])
    \right),
\end{equation}
\begin{equation}
    g^{\mathrm{spf}}_t
    = \sigma\left(
    \operatorname{MLP}_{\mathrm{spf}}
    ([h^{\mathrm{omni}}_t;h^{\mathrm{spf}}_t])
    \right),
\end{equation}
where $g^{\mathrm{omni}}_t,g^{\mathrm{spf}}_t
\in(0,1)^H$. The two gating networks have independent
parameters. The gate $g^{\mathrm{omni}}_t$ controls
the contribution of the Omni features, while
$g^{\mathrm{spf}}_t$ controls the contribution of the
projected spoof features.

The fused feature at position $t$ is computed as
\begin{equation}
    h_t =
    g^{\mathrm{omni}}_t\odot h^{\mathrm{omni}}_t
    + g^{\mathrm{spf}}_t\odot
    \tilde{h}^{\mathrm{spf}}_t
    \in\mathbb{R}^{H}.
\end{equation}
The final representation for the entire utterance is
obtained by concatenating the fused features across
all temporal positions:
\begin{equation}
    h=[h_1;\ldots;h_T]\in\mathbb{R}^{T\times H}.
\end{equation}

To initialize the model with the original Omni
representations, we initialize the final layer of
$\operatorname{MLP}_{\mathrm{spf}}$ to zero and set
the bias of the Omni gate such that
$g^{\mathrm{omni}}_t\approx\mathbf{1}$.
Consequently, the initial fused representation
satisfies $h\approx h^{\mathrm{omni}}$.

Compared with APSGF, this baseline retains the same dual-gate fusion mechanism but removes the ASR-protected subspace constraint, allowing both Omni and spoof features to be adaptively fused across the full Omni feature space. This comparison isolates the effect of subspace protection on joint ASR and word-level spoof detection.

\begin{table}[!t]
\centering
\caption{Parameter counts of the ablation models. Total parameters include all instantiated model parameters, while trainable parameters include the DoRA weights and their associated modules.}
\label{tab:param_counts}
\begin{tabular}{c c c | rr}
\toprule
 \textbf{APSGF} &\textbf{RAPO} & \textbf{Total (M)} & \textbf{Trainable (M)} \\
\midrule
\xmark &\xmark     & 4731.7 & 28.2 \\
\cmark &\xmark  & 5065.5 & 46.6 \\
\xmark &\cmark  & 4731.7 & 28.2 \\
\cmark &\cmark  & 5065.5 & 46.6 \\
\bottomrule
\end{tabular}
\end{table}

\section{GRPO Implementation Details}
\label{grpo}

GRPO is initialized from the epoch-5 SFT checkpoint with a fresh AdamW optimizer state. The learning rate is $5\times10^{-6}$, with a cosine decay schedule and a warmup ratio of 0.03. Training is performed for one epoch.

For each prompt, GRPO samples 8 responses. The sampling temperature is 0.90, with top-$p$ set to 0.95 and top-$k$ set to 50. The maximum generation length is 1024 tokens. GRPO uses the same reward function and KL regularization settings as RAPO, with a KL coefficient of 0.04. The remaining GRPO-specific settings follow the configuration used in our experiments.

\section{Parameter Counts of Ablation Models}
\label{Parameter}
Table~\ref{tab:param_counts} reports the total and trainable parameter counts of the ablation models. The total count includes all instantiated parameters, including the frozen Qwen2.5-Omni Thinker and the additional frozen Wav2Vec2-XLS-R-300M encoder where applicable.  Trainable parameters comprise the DoRA weights and their associated modules. RAPO introduces no additional model parameters. APSGF increases the parameter count by introducing an auxiliary feature extractor and additional gating networks for feature fusion.

\end{document}